\documentstyle[12pt]{article}

\begin{document}
\pagestyle{empty}
\begin{center}
{\LARGE Saturation in the Nuclear Matter Problem}
\\
\vspace*{1cm}
{\sc G. E. Brown}~\footnote{Supported, in part, by the U.S. Department
of Energy Grant No.~DE-FG02-88ER40388; in Germany by the
Humboldt Foundation.}\\
{\it Physics Department,
State University of New York at Stony Brook, New York 11794, and
IKP (Theory) Kernforschungsanlage J\"{u}lich, 5170 J\"{u}lich, Germany,}\\
and\\
{\sc R. Machleidt}~\footnote{Supported  by the U.S. National
Science Foundation under Grant No.~PHY-8911040 and PHY-9211607.}\\
{\it Department of Physics, University of Idaho,
Moscow, Idaho 83843}
\\
\vspace*{1cm}
September 8, 1992
\end{center}

\newpage
\setcounter{page}{1}
\pagestyle{plain}

\begin{abstract}
Once density-dependent meson masses are introduced into the
nuclear many-body problem, conventional mechanisms for
saturation no longer operate.
We suggest that a loop correction, essentially the introduction of the
axial vector coupling $g_A(\rho,k)$ as function of density $\rho$
and momentum $k$, can bring about saturation, and
present schematic calculations to illustrate this.
We find that a very small density-dependence in $g_A$
gives rise to a very large saturating effect on nuclear matter. In fact,
this new saturation mechanism
turns out to be
more powerful than any of the conventional mechanisms.
\end{abstract}


\pagebreak

%\section{Introduction}
For many years during which nonrelativistic Brueckner theory was employed,
saturation could not be achieved at nuclear-matter density in the
nuclear many-body problem~\cite{Day83}.
 The Walecka mean-field theory~\cite{SW86}
appeared to solve this problem. In this theory, the scalar density
$\rho_s$, which coupled to the attractive scalar interactions,
increased less rapidly with density than the vector density $\rho_V$,
which coupled to the repulsive vector interactions. As a result,
saturation occurred, and this could easily be arranged
to happen at nuclear matter density $\rho_0$.

Recent calculations~\cite{BM90,LMB92,BT92}
have shown that rather complete relativistic Brueckner-Hartree-Fock
calculations reproduce very much the same scalar and vector mean fields
from the Bonn potential for the nucleon-nucleon interaction~\cite{MHE87,Mac89}
as have been used in Walecka's mean field theory.
Thus, the mean fields of the Walecka theory can be derived from an
interaction which describes nucleon-nucleon scattering well.
This would seem to bring the story of nuclear saturation to a
successful conclusion.

There are, however, new elements in the problem. In a number of papers
(see, e.~g., Ref.~\cite{BMP90}) it was proposed that meson masses decrease,
 with increasing density, at about the same rate as the nucleon
{\it effective} mass. (The quasiparticle velocity $v_{QP}=p/m^*_N$
defines the effective mass.)
Reasons for the dropping meson masses were most economically summarized
by Brown and Rho~\cite{BR91}.
Rather complete calculations of the density dependence of
the vector-meson masses $m^*_{\rho}$ and $m^*_{\omega}$ have
now been carried out by Hatsuda {\it et al.}~\cite{HL92}
within the QCD sum rule (or finite energy sum rule) formalism.
The authors find
\begin{equation}
\frac{m_V(\rho)}{m_V(0)} \approx 1-C\: \frac{\rho}{\rho_0}
\end{equation}
where $m_V$ is the vector-meson mass, with $C\approx 0.18$.
We believe the error quoted as $\pm 40$\% in Ref.~\cite{HL92}
to be larger than the real uncertainty in the calculation, which can
be cast in a form to depend chiefly on the value of the nucleon
sigma term $\Sigma_{\pi N}$;
which has fairly reliably been pinned down to be
$\approx 45$ MeV~\cite{GLS91}.

We wish to point out that one of the most often employed
nucleon effective masses in nuclear structure calculations
is that of the SkM$^\star$ potential~\cite{Bartel}
\begin{equation}
\frac{m^*_N}{m_N} = \frac{1}{1+0.25\rho/\rho_0}
\end {equation}
which gives $m^*_N/m_N = 0.8$ for $\rho = \rho_0$.
Whereas one can see from the calculations~\cite{HL92}
that the mass of the vector meson $m_V^*$ mainly follows the
cube root of the quark condensate $<\bar{q}q>$, roughly as
\begin{equation}
\frac{m^*_V}{m_V} \approx  \left( \frac{<\bar{q}q>^*}{<\bar{q}q>}
\right)^{1/3}\; ,
\end{equation}
the nucleon effective mass involves not only the
scalar condensate $<\bar{q}q>^*$, but also the vector one
$<q^{\dagger}q>$, and is much more difficult to calculate in the
QCD sum rule formalism~\cite{FGC92}.
Thus, the $m^*_N/m_N$ may not accurately follow that of $m^*_V/m_V$,
but it seems reasonable that these ratios are about the same
at nuclear matter density $\rho_0$; and we shall make the asumption
that they scale together as in Ref.~\cite{BR91}.

The pion mass $m_\pi$ does not change appreciably with density~\cite{BRK91}.
Thus, when a common scaling is assumed for meson masses, we must go
back and make a correction for the lack of scaling of the pion mass.

Given that masses decrease with a common scale
\begin{equation}
\lambda(\rho) \equiv \frac{m^*_N}{m_N}=\frac{m^*_V}{m_V}= ... \; ,
\end{equation}
then this common scale can be taken out~\cite{BMP90}
\begin{equation}
H(r)=\lambda(\rho) H_{vac}(\lambda r) = \lambda H_{vac} (x)
\end{equation}
where $x=\lambda r$ and where $H_{vac}$ is the original Hamiltonian
for the many-body system with
$r$ replaced by $\lambda r$, but with vacuum masses for
both nucleon and mesons.
 From Eq.~(5), which we assume to hold at least approximately, we
learn two important results:
\begin{enumerate}
\item There is no saturation, with the assumption Eq.~(4), at this level.
(Although fairly obvious, this will be demonstrated below by
 explicit calculations.)
\item Given the scaling Eq.~(4), the net binding energy of nuclear matter
depends only linearly on the effective mass,
proportional to $m^*_N/m_N$,
a very weak dependence, slightly favoring larger $m^*_N$ for greater binding
energies. But we will scarcely be able to calculate the energy $E$
accurately enough in order to determine $m^*_N$ from such a calculation.
\end{enumerate}

Certainly, the lack of saturation in our theory with scaling masses is a grave
deficiency. Brown {\it et al.}~\cite{BMP90} forced saturation in an
{\it ad hoc} fashion by including a two-loop correction in the
phenomenological theory, but this is unsatisfactory in the long run.
 Since this work, it has been realized~\cite{BR91} that the scaling
operates at best only at the mean field level,
mirroring changes in the vacuum condensate. Loop corrections go beyond
the scaling Eq.~(4) and introduce new scales, at least in
certain ranges of density. The most familiar loop
correction in nuclear physics is that which changes the ratio
$g_A/g_V$ from unity at the level of tree diagrams (in, e.~g., the
linear $\sigma$-model) to 1.26 at zero density.
It can be calculated by the Adler-Weisberger relation.
We shall show now that the density dependence
of this loop correction can provide a mechanism for saturation.

%\section{Discussion of the Loop Correction}

Whereas $g_A=1.26$ at zero density, there are many empirical indications
that it decreases with density. Both Rho~\cite{Rho74} and
Ohta and Wahamatsu~\cite{OW75} find
\begin{eqnarray}
\frac{g_A(\rho)}{g_A(0)} & = & \left[ 1 + \frac89 \left(
\frac{f_{\pi N \Delta}}
{m_\pi} \right)^2 \rho \: \frac{(g_0')_{N\Delta}}{\omega_R} \right]^{-1}
\\
   & = & \left[ 1 + b \: \frac{\rho}{\rho_0} \right]^{-1}
\end{eqnarray}
with $b \approx 0.8 (g_0')_{N \Delta}$,
where $f_{\pi N \Delta}$ is the pion coupling between nucleon and
$\Delta$ ($f_{\pi N \Delta} \approx 2$), and $\omega_R$ is the energy
difference between $\Delta$ and nucleon ($\omega_R \approx 300$ MeV).

There is currently controversy as to whether $g_A(\rho)$ drops
this rapidly with increasing density, but it is clear that in many experiments
Gamow-Teller strength is missing~\cite{Goo86}, and this indicates that $g_A$ in
nuclei is less than $g_A(0)$.

As noted in the last section, if we begin from the pion coupling at
tree approximation in the linear $\sigma$ model (which corresponds
to the mean field approximation of Brown and Rho~\cite{BR91}), then
\begin{equation}
g_{\pi NN} = (g_{\pi NN})_{tree} \: g_A \; .
\end{equation}
In the linear $\sigma$ model, the scalar $\sigma$ meson is united,
by chiral invariance, with the pion in a four-vector. Although, the scalar
particle with $m_S \approx 550$ MeV used in boson-exchange models is
fictious, summarizing some enhancement in the correlated two-pion
exchange, we shall none the less find it useful to treat it
symmetrically with the pion, for reasons we explain below.
We thus give this fictious
scalar particle also a loop correction,
\begin{equation}
g_{SNN} = (g_{SNN})_{tree} \: g_A \; .
\end{equation}

The precise rate predicted for the decrease, with density, for $g_A(\rho)$
in Eq.~(6) depends on the
value of $(g_0')_{N \Delta}$, which seems to be in the region~\cite{Joh88}
\begin{equation}
0.4
\; \raisebox{-.3ex}{\small $ \stackrel{\textstyle <}{\sim} $} \;
(g_0')_{N \Delta}
\; \raisebox{-.3ex}{\small $ \stackrel{\textstyle <}{\sim} $} \;
0.5 \; .
\end{equation}

Whatever the precise value of $(g_0')_{N\Delta}$, $g_A$ is predicted
to drop from 1.26 to close to unity as $\rho$ increases from zero to
$\rho_0$.
This would mean a large (gigantic) decrease in scalar strength in the
nucleon-nucleon interaction, which would obviously upset whatever
present agreement there is between calculations and nature.

In order to incorporate the loop correction without upsetting
completely previous calculations, we must generalize $g_A$ to be
a function of both density and momentum
\begin{equation}
g_A \longmapsto g_A(\rho,k) \; .
\end{equation}
As we next show, there is considerable cancellation between the density and
momentum dependencies, but enough dependence on density will be left
for it to be important to include $g_A$.

The two-pion-exchange contribution in the nucleon-nucleon interaction
gives the lowest-mass effective scalar exchange.
Let us consider the density dependence in this to lowest order in $\rho$;
see Fig.~1. Such medium corrections, as well as the $\pi-\pi$ rescattering have
recently been calculated in considerable detail by
Mull, Wambach, and Speth~\cite{MWS92}.
Our simple calculations here should be viewed as a schematization of
the complete calculation, with which they agree not only qualitatively,
but semiquantitatively.

The formula Eq.~(6) is obtained by retaining only the process of Fig.~1b for
momentum $k=0$, but summing the bubble contribution to all
orders. The renormalized operator $g_A(\rho)$ is then to be used in
the nucleon space, and the process of Fig.~1a will be included by
configuration mixing. We find it more convenient
to include both processes Figs.~1a and 1b,
in principle summed to all orders, in our definition of
$g_A(\rho,k)$.

It is clear that the momentum dependence of $g_A$ cancels most of
the density dependence, as we now show. The modification in the
pion propagator due to the isobar-hole insertion can, to linear order in the
density, be expressed through the factor~\cite{Ain88}
\begin{equation}
\delta g_A(\rho,k) = 1+ 0.8 \: \frac{\rho}{\rho_0} \left[
\frac{k^2}{k^2+m_\pi^2} - (g_0')_{N\Delta} \right]
\end{equation}
where we have left out effects of retardation in the pion propagator.
Reference~\cite{Ain88} takes the typical pion loop momentum to be
\begin{equation}
<k^2> \approx 3 m_\pi^2 \; ,
\end{equation}
so that the first factor $<k^2/(k^2+m_\pi^2)> \approx 3/4$ would
seem to predominate over $(g_0')_{N\Delta}$, giving
an increase in $g_A(\rho,k)$. More detailed calculations~\cite{MWS92,BLL92}
which include the momentum dependence of $(g_0')_{N\Delta}$ show the
latter quantity
to grow sufficiently with momentum so as to be comparable or slightly
greater than 3/4 at the momentum given by Eq.~(13).
It is clear that there is a close cancellation between
the $k^2/(k^2+m_\pi^2)$ and $(g_0')_{N\Delta}$ terms in Eq.~(12).

In the case of the nucleon particle-hole insertion, the $(g_0'(k))_{NN}$
is found to be $\approx 15$\% larger than the $<k^2/(k^2+m_\pi^2)>
\approx 3/4$ term. (Both are weighted by a common Lindhard function
in this case.)
The net result is that $g_A(\rho,k)$, as we define it, is decreased
from $g_A(0,0)$ by a factor such as is shown in Eq.~(7), but with $b$
replaced by
\begin{equation}
b_{eff} \approx \frac{1}{10} b
\end{equation}
In other words, the near cancellation between repulsive and
attractive effects, the latter from the momentum dependence, cancel most
of the decrease with density.
In this way, we can understand why investigators have not found
it necessary to include the loop correction $g_A$ to date.
The fact that there is this large cancellation, leaving a small
repulsive effect for the scalar
strength, is born out by the detailed calculation of Mull {\it et al.}
\cite{MWS92}.

%\section{Calculations}

In our nuclear matter calculations,
we start from the Bonn A potential~\cite{Mac89} and employ the relativistic
density-dependent Brueckner-Hartree-Fock formalism~\cite{BM90,Mac89}.

In the bottom of Fig.~2, we show the energy of nuclear matter as
function of $k_F$ for two cases. In Case~1, the nucleon effective mass
$m^*_N$ and meson masses $m^*_m$ of the Bonn potential,
other than $m_\pi$ (which is held fixed) are taken to scale as in
Ref.~\cite{BMP90} (Table~1, last column, therein).
 In Case~2 the nucleon effective mass is taken to scale as in Eq.~(2), and the
meson masses, other than that of the pion,
scale in the same way. The behavior of the curves can be understood from our
discussion following Eq.~(5).

The upper curves in Fig.~2,
which exhibit saturation,
are obtained by multiplying $g_{SNN}$ and $g_{\pi NN}$ by
\begin{equation}
(g_A)_{eff} = \frac{1}{1+b_{eff} \rho/\rho_0}
\end{equation}
In Case~1, $b_{eff} = 0.035$ and in Case~2, $b_{eff} = 0.045$.
These are about an order of magnitude less than the $b\approx (g_0')_{N\Delta}$
of Eq.~(7), for the reasons noted. Obviously, it will be difficult to
calculate $b_{eff}$ with any precision.

By comparing the upper curves in Fig.~2 with the lower ones, it is clearly
seen that even this very small density-dependence of $g_A$ has a very large
saturating effect on nuclear matter. In fact, this effect is
even larger than the Pauli effect that is known to be the most
powerful one among  conventional saturation mechanisms.

The saturation energy in both cases shown in Fig.~2 is --15.3 MeV
per nucleon. The saturation density corresponds to $k_F=1.35$ fm$^{-1}$
for Case~1 and $k_F=1.30$ fm$^{-1}$ for Case~2. These saturation densities are
better than those from relativistic Brueckner-Hartree-Fock calculations
where  $k_F=1.40$ fm$^{-1}$ is obtained together with an energy
of --15.6 MeV per nucleon~\cite{BM90}.

%\section{Discussion}

In the QCD sum rule calculations~\cite{DL90,DR92,CFG91}
the lowest order contribution, which gives the main part
of the scalar and vector fields, is linear in density,
just as are the mean fields in the Walecka theory. In this sense, the
Walecka mean field theory rather directly mirrors QCD~\cite{CFG91}.
Our work supports this.
The way this comes about in detail, beginning from two-body interactions
 which reproduce the nuclear scattering, such as the Bonn
potential~\cite{MHE87,Mac89},
is, however, intricate. In Refs.~\cite{BM90,LMB92,BT92}
it can be seen that the effective scalar and vector mean fields arising
from the Bonn potential evaluated in the relativistic
Brueckner-Hartree-Fock (RBHF) calculation are equal to the Walecka
mean fields only in the vicinity of nuclear matter density.
The effective coupling constants in RBHF are density dependent,
after the effects of short-range correlations and exchange are
taken into account. For example, at $k_F=0.8, 1.1, 1.5$ fm$^{-1}$
Ref.~\cite{BT92}
obtains $g^2_\sigma/4\pi=12.3, 8.91, 6.23$ and
$g^2_\omega/4\pi=18.63, 13.48,9.06$,
respectively, in agreement with Ref.~\cite{BMP90}.
At normal nuclear matter density, $k_F=1.35$ fm$^{-1}$, the scalar
field is $S=-355.7$ MeV, the vector field $V=+275$ MeV~\cite{BM90},
very close to the values normally employed in the Walecka theory.

For completeness
we should note that Drukarev and Levin~\cite{DL91}
have calculated the density dependence of $g_A$; i.~e.,
$g_A(\rho)$, from QCD sum rules. They essentially confirm the results
of Refs.~\cite{Rho74,OW75}. We suggest that it may be necessary to
include the effects of $g_A(\rho,k)$ in the QCD sum rule calculations
of the binding energy in order to achieve saturation.

One of the authors (G.E.B.) would like to thank J. Speth and J. Wambach
for extremely useful discussions.

\pagebreak
\begin{center}
\large Figure Captions
\end{center}

{\bf Figure 1.} Corrections, to linear order, for the density in
two-pion exchange. In a) the nucleon particle-hole insertion is
shown; in b) the isobar-hole insertion.
In both cases, a linear density-dependence results from the sum over hole
states. The cross-hatched insertions in the bubbles represent
the vertex corrections $(g_0')_{NN}$ and $(g_0')_{N\Delta}$,
respectively.

{\bf Figure 2.} Energy per nucleon in nuclear matter.
The bottom lines (dashed) represent two energies,
as function of $k_F$, with scaling masses.
Curve~1 refers to Case~1 and curve~2 to Case~2 (cf. text).
The correction $<g_A(\rho)>$ gives the solid curves in the top of the figure.
 The dotted line is
the conventional relativistic Brueckner-Hartree-Fock (RBHF)
result~\cite{BM90}.
(In all cases, then Bonn A potential~\cite{Mac89} is applied.)
The cross denotes the empirical value for nuclear matter saturation.

\pagebreak

\vspace*{4cm}

The figures can be obtained upon request from R. Machleidt
(machleid@idui1.bitnet; machleid@tamaluit.phys.uidaho.edu).
Please specify if you want them by ordinary mail or telefax and include
your mailing address or fax number with your request.

\end{document}